\documentclass[conference]{IEEEtran}
\usepackage{graphicx}
\usepackage{color}

\begin{document}
%
\title{DiscoverFriends: Secure Social Network Communication in Mobile Ad Hoc Networks}

\author{\IEEEauthorblockN{Joshua Joy*}
\IEEEauthorblockA{UCLA\\
jjoy@cs.ucla.edu}
\and
\IEEEauthorblockN{Eric Chung*}
\IEEEauthorblockA{UCLA\\
chung.yanhon@gmail.com}
\and
\IEEEauthorblockN{Zengwen Yuan}
\IEEEauthorblockA{UCLA\\
zyuan@cs.ucla.edu}
\and
\IEEEauthorblockN{Leqi Zou}
\IEEEauthorblockA{UCLA\\
zlqiszlq@cs.ucla.edu}
\and
\IEEEauthorblockN{Jiayao Li}
\IEEEauthorblockA{UCLA\\
likayo@ucla.edu}
\and
\IEEEauthorblockN{Mario Gerla}
\IEEEauthorblockA{UCLA\\
gerla@cs.ucla.edu}}


%


\IEEEoverridecommandlockouts

\maketitle

\let\thefootnote\relax\footnotetext{*Eric and Joshua contributed equally to this work.}


%
\IEEEpeerreviewmaketitle

\begin{abstract}

This paper presents a secure communication application called DiscoverFriends.  Its purpose is to securely communicate to a group of online friends while bypassing their respective social networking servers under a mobile ad hoc network environment. DiscoverFriends leverages Bloom filters and a hybrid encryption technique with a self-organized public-key management scheme to securely identify friends and provide authentication. Additionally, DiscoverFriends enables anonymous location check-ins by utilizing a new cryptographic primitive called Function Secret Sharing. Finally, to the best of our knowledge, DiscoverFriends implements and evaluates the first Android multi-hop WiFi direct protocol using IPv6. 


\end{abstract}

\section{Introduction}

Mobile devices (e.g. smartphones and tablets) have become a rising
platform for Online Social Networks (OSNs). Facebook, for example,
reported over 1 billion mobile users \cite{statista} in the first
quarter of 2014. Why are so many users moving to the mobile platform?
One possible reason is that users can have instantaneous access to
OSN with mobile devices at any time and place. But more importantly,
users are attracted by the various interesting features that are only
available on mobile platforms. Location-based services (LBS) \cite{i._a._junglas}
is one of the top features exploited by mobile users.

With GPS integrated into smartphones, mobile OSN users can easily
localize themselves, which allows them to share their own location
and related experiences with friends (e.g., the ``Check-in'' feature
of the Facebook mobile app). Since it does so, users are more able
to find nearby friends based on their locations. Therefore, the LBS
functionality provides great convenience for users' social activities.

However, the application of LBS in OSNs also incurs some concerns
surrounding the issue of privacy. When a user activates the LBS function
in an OSN application, it exposes the user's location to the OSN service
provider. In other words, Facebook can track the user's activity with
this location information. One question is: Can the OSN users still
utilize partial LBS without reporting their location to OSN providers.
Typically, can we find nearby friends without requesting them from
the Facebook servers. This paper provides a solution to this problem.

In this work, a Wi-Fi-based solution called DiscoverFriends helps
an OSN user find nearby friends. The core idea is to leverage the
local Wi-Fi communication to directly discover an OSN friend while
bypassing OSN servers. The design faces two main challenges. The first
problem is how to authenticate an OSN user without going to the OSN
server. Second, due to the broadcast nature of wireless communication,
the messages can be overheard by other users. The potential eavesdropping
can be a new threat to the privacy. To this end, DiscoverFriends addresses
these issues using a Bloom filter-based approach with hybrid encryption
to provide a higher level of security.

The solution provided by DiscoverFriends is not only limited to discovering
nearby friends. It can further be employed to setup communication
between friends in the same local Wi-Fi. In other words, OSN users
can exchange multi-hop text messages and other data without going through the
OSN server, which further helps to prevent users from being tracked
by those OSN providers.

Finally, DiscoverFriends provides users a mechanism to anonymously "Check-in" their location on an OSN. To hide their identity, a user could "Check-in" and broadcast their location under a pseudoynm (shared previously with friends). However, the OSNs can easily examine the authenticated connection and infer the true identity. Discoverfriends utilizes a new cryptographic primitive called Function Secret Sharing (FSS) that is robust against traffic analysis attacks. As long as there are more than two users announcing their location, it is not possible to infer which user posted a particular "Check-in". FSS scales well to large anonymity sets as FSS is asynchronous and uses an efficient PRG.

This paper is organized as follows. Section 2 begins by elaborating
on the application design followed by implementation details presented
in Section 3. Section 4 describes the risk analysis under two threat
models. Performance evaluation in Section 5 is done against alternative
schemes described in the related works in Section 6. Lastly, Section
7 summarizes the features of the application and concludes the paper. 

\section{Design}

In DiscoverFriends, any OSN user can serve as an $Initiator$ and
leverage Wi-Fi broadcast to send request messages in order to find
a specified friend, hereby referred to as $Target$. If the target
receives the request, the target will then send back an acknowledgement
to the initiator. The request message is constructed using Bloom filters
and a certificate to achieve confidentiality and authentication, respectively.

The solution proposed is built based on an important assumption: each
OSN user has a confidential ID, which is only accessible to the user's
friends. This ID is not necessarily the string that OSNs use to identify
each user, but it can be an extra confidential string. For instance,
Facebook provides a user with a public username and private ID. This
assumption is reasonable since many OSNs have access control mechanisms,
in which users can specify the accessibility of private information
(e.g., whether it is public to all users or friends only).

Furthermore, DiscoverFriends is not limited to supporting only one
OSN as it can utilize IDs from multiple OSNs to improve security.
More specifically, users can XOR their IDs in Facebook and Google+
to generate a new ID. This new ID cannot be recognized by any single
OSN provider even if the provider somehow captures the message. One
drawback of this technique is that the intended friend has to be the
initiator's friend on multiple OSNs, which may limit the usage of
the application.

In this section, the application of Bloom filters is introduced and
then explained how they are applied in DiscoverFriends. In addition
to that, two encryption mechanisms are analyzed and a combined solution
is selected to be used in the application. Moreover, an overview of
key management techniques is provided, where one scheme is selected
to be used in the application. Next, Function Secret Sharing (FSS) is introduced and explained how anonymous check-ins are achieved. Finally, the communication protocol is examined between the initiator and the target.

\subsection{Bloom Filter}

Bloom filter is a space-efficient probabilistic data structure, whose
purpose is to test whether a specific element is in a given set or
not. It may produce false positive results but will definitely not
return false negative results. Bloom filter is widely used in caching
mechanisms as well as security solutions.

\begin{figure}[tb]
\includegraphics[width=1\columnwidth]{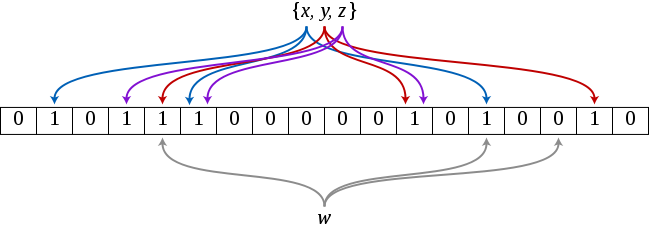}

\protect\caption{Bloom filter representation where \textit{x,y,z} represent the elements in the set. }
\label{bloom_filter_representation}
\end{figure}

Bloom filter is simply an $m$-bit array. When an element is added
into a Bloom filter, $k$ hash function is applied on the element
to get $k$ hash values in the range of {[}0, $m$). Then, the corresponding
bits in the array are set to 1. When testing for an element, the same
set of hash functions are applied before checking to see whether the
corresponding bits in the Bloom filter are set to 1. If all of them
are 1, then it means that the given element is included in the set;
otherwise, it is excluded in the set. This explanation can be shown
in Figure \ref{bloom_filter_representation}.

When designing the Bloom filter, one focus is to optimize the number
of hash functions $k$ and the length of the Bloom filter $m$, which
can be calculated as a function of $n$, the number of inserted items,
and $p$, the desired false positive probability as shown in Equation
\ref{eq:m}. In the case of DiscoverFriends, $n$ is the number of
targeted friends the initiator wants to communicate to. Also, the
choice of a small $p$ value is necessary as it corresponds to the
probability that an attacker may be able to ascertain the hash of
the initiator's ID. Subsequently, the appropriate $k$ value can be
computed using $m$ and $n$ using Equation \ref{eq:k} \cite{DBLP:journals/cacm/Bloom70}.

\begin{equation}
m=-\frac{n\ln p}{\left(\ln2\right)^{2}}\label{eq:m}
\end{equation}

\begin{equation}
k=\frac{m}{n}\ln2\label{eq:k}
\end{equation}

In DiscoverFriends, Bloom filters serve as the container of the IDs
for friends. More specifically, when the initiator, who will be called
Alice, sends the initial request, she adds the target's ID into a
Bloom filter, which will be sent during the network initialization
phase for friend discovery. Thus, when the target, say Bob, gets the
request, he can determine whether or not the request is directed to
him. In addition, Alice will also include her own ID in another Bloom
filter carried in the request, so the target can also figure out who
the initiator is. The network initialization phase does not need to be done over a secure channel as the Bloom filters provide the required level of security.

By using this technique, OSN users are not required to put their confidential
ID into the broadcast message, which can be listened to by other people.
Therefore, even if an adversary, Eve, eavesdrops the request message,
it is extremely difficult for her to infer the initiator's ID if she
is not the initiator's friend because the adversary only knows the
hash values and cannot acquire the accurate corresponding ID with
the hash values, let alone understand the user corresponding to that
ID. However, it is important to note that although the use of Bloom
filters can speed up initiator ID discovery by using fast hash algorithms to encrypt the confidential IDs and reduce latency by passing the lightweight Bloom filters over the network, it
makes it easier for non-friends to discover the ID. Thus, security
is traded off at the expense of lower discovery latency.

\subsection{Digital Signature and Key Management}

To ensure the security of the application, the idea was to use certificates
for authentication. In order to generate a public key certificate,
two schemes exist: public key infrastructure (PKI) and web of trust.
In PKI systems, a certificate authority (CA) issues certificates that
binds public keys to identities, and this information is kept in a
central repository. On the other hand, a web of trust entails identity-based
cryptography through self-signed certificates, where a key generation
center (KGC) generates users' private keys. Evidently, this scheme
is highly susceptible to man-in-the-middle attacks.

However, things are not that simple in an infrastructure-less model.
Because of the ad hoc nature of DiscoverFriends, conventional public
key infrastructure and web of trust schemes are less suitable for
the application due to complicated certificate management \cite{k._sadasivam}.
Also, certificate authentication becomes impractical. The problem becomes
clear when nodes cannot guarantee their connectivity and cannot rely
on infrastructure to detect compromised mobile nodes \cite{j._macker}.
In addition, certificate revocation in MANETs \cite{p._rathi} becomes
an issue because of the lack of centralized repositories and trusted
authorities.

There are two main types of public-key cryptography: certificate-based
cryptography (CBC) and ID-based cryptography (IBC). Previous works
\cite{s._yi} have attempted to use CBC in MANETs. A naive approach
for key distribution occurs in the network setup phase, where every
node is preloaded with all other's public key certificates. However,
as mentioned above, key management becomes an issue as key updates
need to be handled in a cost-effective manner \cite{j.-j._haas}.
As a result, an improvement \cite{s._capkun} using self-organized
public-key management was built on the idea of a certificate chain.
In a different angle, IBC eliminates the need for public-key distribution
and certificates altogether by using a public key based off a public
string identifying an individual. The underlying idea of IBC-based
certificateless public-key management schemes is to have a set of
network nodes share a master-key generated by threshold cryptography
and collaboratively issue ID-based private keys. Built on top of the
underlying concept, variations on the designing of certificateless
public keys have come about in recent literature \cite{y._zhang_1,y._zhang_2,z._zhang}.
The downside of IBC is if the threshold number of nodes gets compromised,
the network is breached.

DiscoverFriends uses a self-organized public-key management approach.
Further details can be read up on in \cite{k._sahadevaiah}. Every
node in the network contains the following:
\begin{itemize}
\item \textbf{Key repository (KR)}: Stores public keys sent by neighbor
node.
\item \textbf{Shared key repository (SKR)}: Stores public keys of all nodes
from the KR.
\item \textbf{Certificate repository (CR)}: Stores valid self-signed certificates.
\end{itemize}
In order to assess the security in the formation of trust, the network
initialization phase plays an important role. The main objective of
this phase is to distribute all the public keys to every node in the
network. Instead of a certificate graph, a trust graph (TG) is generated
by the shared key repository per node. At the end of the initialization
phase, this trust graph is stored as a master graph (MG), which facilitates
frequent public key updates.

\begin{figure}[tb]
\includegraphics[width=1\columnwidth]{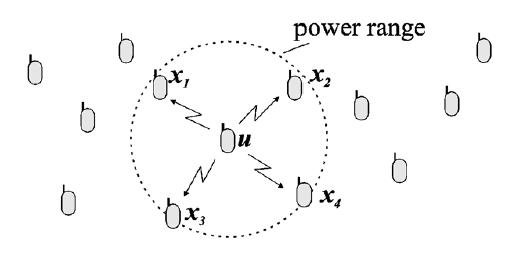}

\protect\caption{Architecture}
\label{architecture}
\end{figure}

\begin{figure}[tb]
\includegraphics[width=1\columnwidth]{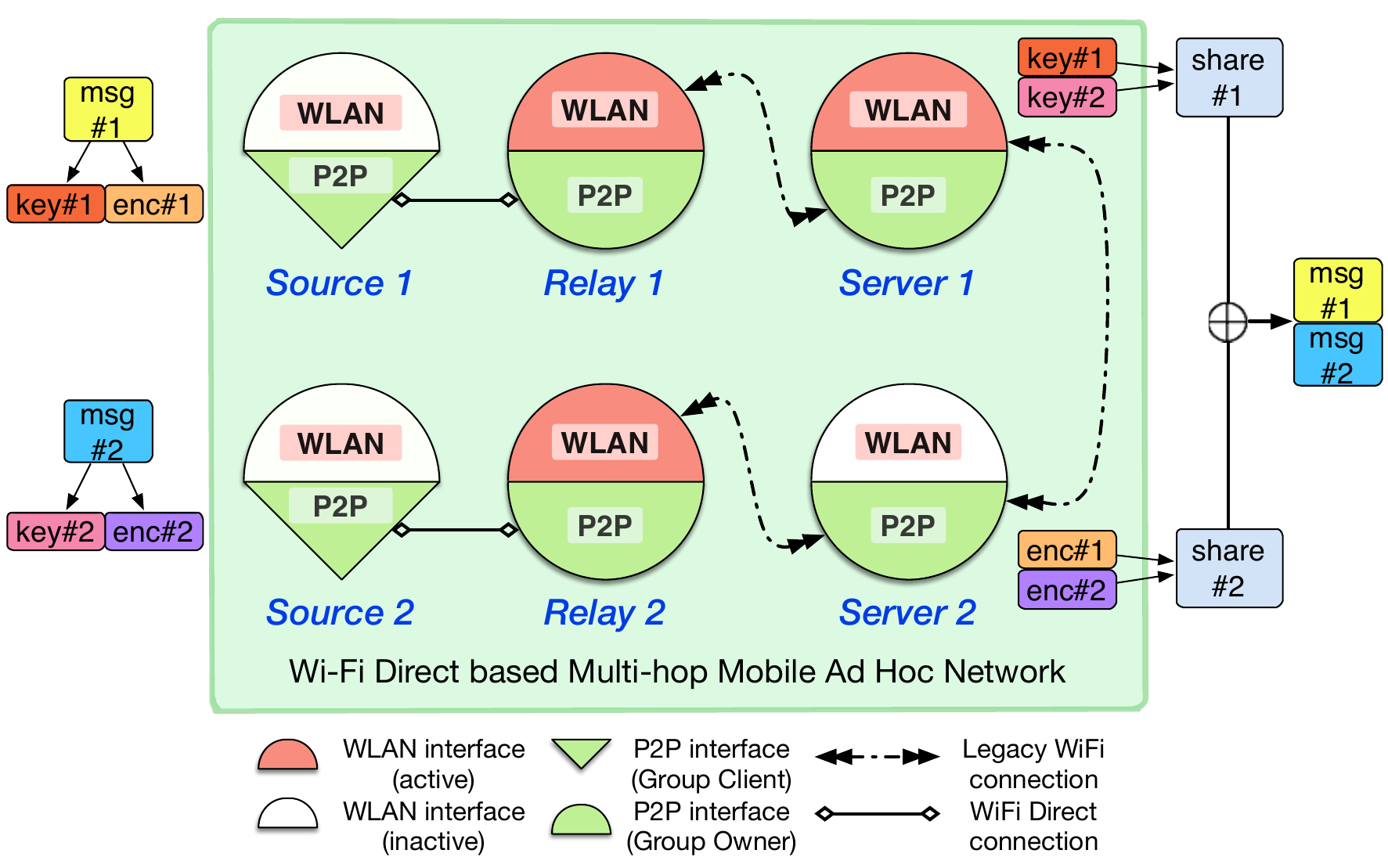}

\protect\caption{Multi-hop Wi-Fi Direct routing}
\label{wifi_direct_routing}
\end{figure}

In its reduced form, DiscoverFriends' model can be simplified as target
nodes do not serve as intermediary nodes for other target nodes as
shown in Figure \ref{architecture}. However, during prolonged application
usage, it may be energy-efficient \cite{d._camps-mur,s._trifunovic}
to pass group ownership to a target node (i.e. x1) once the secure
environment has been set up. This sets the need for an efficient key
management scheme specifically designed for MANETs. In this case,
the target node must know the certificate and public key associated
with all nodes within its group. Therefore, the TG provides a path
to the end node, which has the corresponding private key to decrypt
the sender's message. Within the data structures of each node, valid
certificates and public keys of all other nodes in the group are contained.
The transmission of the information is done during the network initialization
phase, which in the case of DiscoverFriends, happens between friend
discovery and communicating to the group of nodes.

\subsection{Anonymous Check-in}

\begin{figure}[tb]
\includegraphics[width=1\columnwidth]{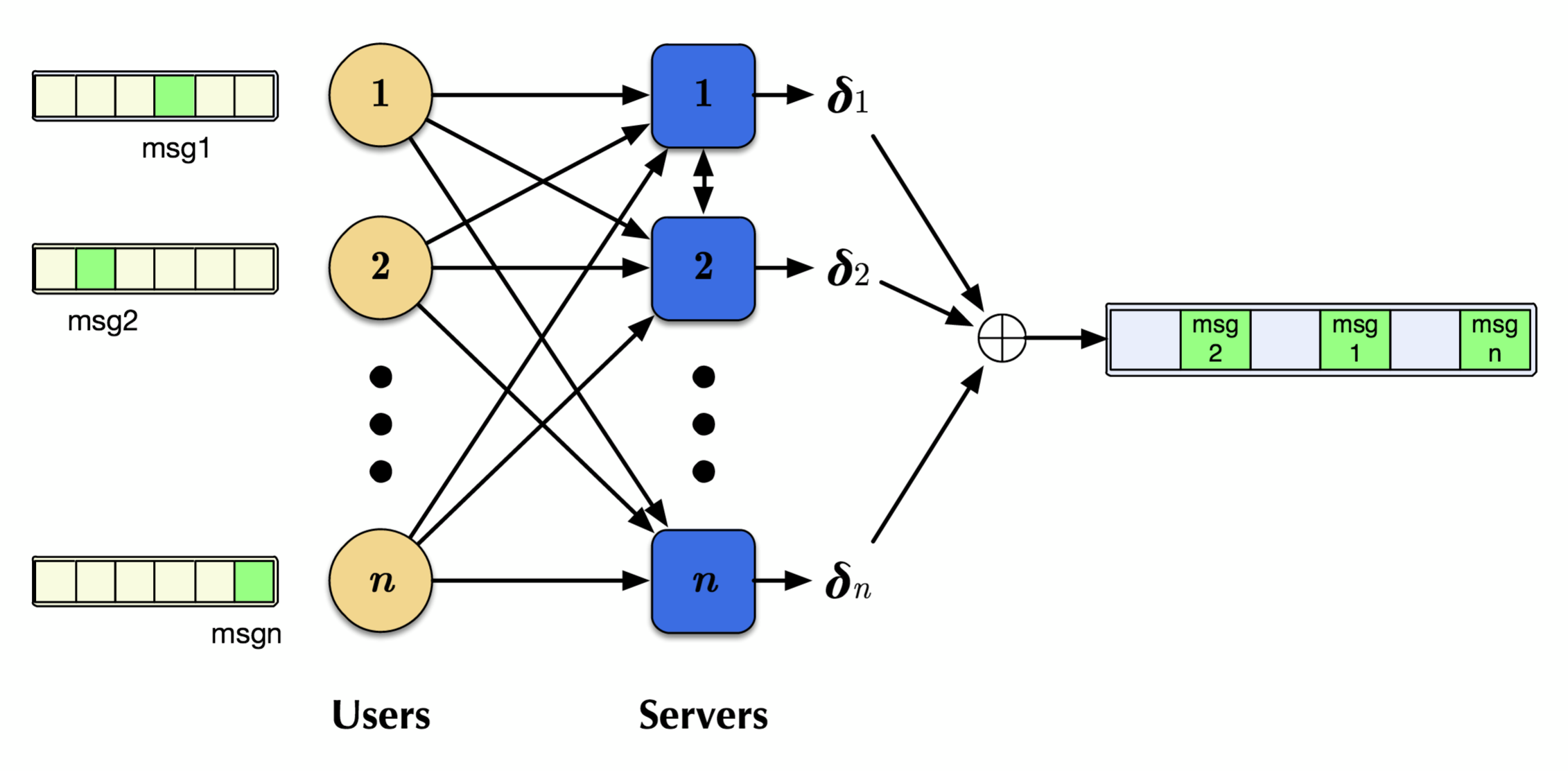}

\protect\caption{Function Secret Sharing}
\label{fss}
\end{figure}

Our goal for anonymity is that a user is able to transmit a message such that the message is unable to be linked back to the user . We assume that each server is operated by a separate OSN and there is at least one honest OSN. 

We utilize Function Secret Sharing (FSS) \cite{DBLP:conf/eurocrypt/BoyleGI15} to anonymously transmit a message that is protected against traffic analysis as long as there is at least one honest server. The intuition behind FSS is as follows. Suppose there is a database with $2^n$ indexes. Each user is able to write their respective message to a randomly selected database index. If each user randomly selects an index and writes to the randomly chosen index, an adversary will not be able to infer which database index a particular user has written to. Thus, FSS defends against traffic analysis.  As long as there is one honest server, the remaining servers will not be able to combine their shares to reconstruct the original message. Each user XORs their message to generate the shares such that the XOR sufficiently randomizes the bits.

We now explain FSS in further detail. Suppose we wish to secretly share a function with \textit{p} parties where at least one party is honest. Suppose there is an input \textit{x} which is \textit{n} bits and the output \textit{y} which is \textit{m} bits. Given \textit{p} keys such that the strings are randomly sampled from the space of ${\{0,1\}}^{2^n*m}$ (total number of inputs multiplied by the size of the message), these strings should evaluate to the message \textit{m} whereby $f(x)=y$ such that $\bigoplus_{i=j}^p k_i[x]=y$. Thus, in this case $p-1$ parties are unable to XOR their keys to discover $f(x)$.

As long as two or more users do not choose the same input x (we assume no collisions), each user is able to write their respective message y to input x. Each user proceeds by sending their keys to each respective server. Each server then performs a bitwise XOR of the evaluation of every $f(x_k)$ over the received key $k$ such that $\bigoplus_{x=0}^{2^{n}-1} f(x_k)$. That is there is a total of $2^n$ evaluations at each server for each key. Each result of $f(x_k)$ is XORed locally at each server resulting in an intermediate computation. This results in an intermediate computation which at the end of the agreed epoch, each server then shares with each other.  \linebreak[3] \newline

\textbf{Intermediate Results}:

\begin{equation}
\delta~shares~server_1= \bigoplus_{x=0}^{2^{n}-1} f(x_{k_1}^a) \oplus f(x_{k_1}^b) ... \oplus f(x_{k_1}^p)
\end{equation}

\begin{equation}
\delta~shares~server_2= \bigoplus_{x=0}^{2^{n}-1} f(x_{k_2}^a) \oplus f(x_{k_2}^b) ... \oplus f(x_{k_2}^p)
\end{equation}

\textbf{Final Output}:

\begin{equation}
server_1=\delta server_1 \oplus \delta server_2
\end{equation}

\begin{equation}
server_2=\delta server_2 \oplus \delta server_1
\end{equation}

\textbf{Servers' Output Should Match}
\begin{equation}
server_1=server_2=\{y_0,y_1,...y_{2^{n}-1}\}
\end{equation}

For our purposes, the particular function we are interested in is a distributed point function. A distributed point function (DPF) maps an input \textit{x} to output \textit{y} such that f(x)=y and f(x')=0 for all x'!=x. Function secret sharing for \textit{p} parties allows to secretly share a DPF amount \textit{p} parties such that the function is unable to be reconstructed with p-1 keys and is thus collusion resistant as long as there is one honest party.

The FSS algorithm works as follows. We consider the special input \textit{x} as a grid where the rows are indexed by the first bits and the columns are indexed by the second bits. The two dimensional grid allows a reduction in the key length to achieve $O(2^{n/2} \cdot 2^{n/2} \cdot m)$.

Next, the function must be cryptographically hidden. To achieve this, $2^{p-1}$ random seeds are chosen for each row. Additionally, $2^{p-1}$ correction words are chosen such that for the special row (the first bits of the  input \textit{x}) XOR together equal the message as the specific point and zeros elsewhere.

The end result should be that the seeds for each row will cancel each other when XORed together, leaving only the special row. To achieve this, special binary arrays are formulated to be multiplied by the seeds to ensure that an even number of the seeds appears in order to cancel each other out except for the special row. To construct the binary arrays, for all rows (except the special row) an even bit array is formulated such that there are \textit{p}-bit columns such that the number of 1 bits is even. For the special row an odd bit array is formulated such that there are \textit{p}-bit columns such that the number of 1 bits is odd.

\subsection{Communication Protocol}

The communication protocol consists of two parties: $Initiator$ and
$Target$. The first two stages represent the network initialization
phase. In the first stage, the initiator, who wants to discover a
certain friend (target), sends a Wi-Fi broadcast request consisting
of three parts:
\begin{itemize}
\item \textbf{$BF_{c}$}: The Bloom filter containing the $Target$'s ID.
\item \textbf{$BF_{c}$}+: $BF_{c}\oplus hash\, of\, Initiator's\, ID$.
\item \textbf{$CF$}: $Initiator$'s certificate encrypted with AES; the
encryption key is the hash of the $Initiator$'s ID.
\end{itemize}
Here, $BF_{c}$ is used to identify whether the target is a person
of interest in the invitation. $BF_{c}+$ builds on top of this knowledge
and helps the target identify the initiator. Lastly, $CF$ is used
for authentication without going through the OSN server. In detail,
the certificate contained inside is self-signed by the initiator.

In the second stage of the protocol, the target receives the request
message and proceeds in the following steps:
\begin{enumerate}
\item Test whether the $Target$'s own ID is in $BF_{c}$ . If not, terminate
the procedure. Otherwise, proceed to next step.
\item $BF_{c}\oplus BF_{c}+\rightarrow hash\, of\, Initiator's\, ID$.
\item Traverse the $Target$'s own friend list and apply hash functions to
each friend's ID to determine if a match with the $Initiator$'s ID
exists.
\item Once a match has been found, take the hash of $Initiator$'s ID to
decrypt $CF$.
\item Check the validity of the certificate in $CF$ and finish authentication.
\item If the $Target$ accepts the invitation from the $Initiator$, the
$Target$ replies with own certificate encrypted with AES; the encryption
key is the hash of the $Initiator$ 's ID.
\end{enumerate}
In stage three, once all peers are connected, the $Initiator$ sends
the set of certificates to the connected peers. Step 6 and stage three
are necessary for the $Initiator$ to update the SKR and CR of other
targets that are connected to the $Initiator$ such that after the
network initialization phase, proper key management is ensured and
eavesdropping attempts are averted. Assuming all targets who want
to connect to the $Initiator$ are connected, the network initialization
phase ends. This leads into stage four and stage five, which represent
subsequent communication. Optionally, step 6 is repeated as many times
as necessary during network connection in order to push certificate
updates to the group as certificates may expire during the communication
session. These two stages may be repeated multiple times until the
connection is broken. In stage four, the $Initiator$ communicates
to the $Target$ as follows:
\begin{enumerate}
\item Generates a random symmetric encryption key.
\item Uses symmetric key to encrypt a message.
\item Encrypts symmetric key using target's public key.
\item Broadcasts the encrypted message and encrypted key.
\end{enumerate}
In response, the $Target$ replies in the following manner:
\begin{enumerate}
\item Uses own private key to decrypt the symmetric key.
\item Uses symmetric key to decrypt the message.
\item Sends back an AES-encrypted message to $Initiator$.
\end{enumerate}
Now, the initiator and target are able to communicate directly, effectively
bypassing OSN servers. 




\section{Implementation}

In this section, details of the implementation are provided, including
the Bloom filter, Wi-Fi communication, and OSN adaptors. 






\textbf{Bloom Filter} One important issue of implementing Bloom filters is the usage of the hash function. There are several candidates, including murmur, fvn series of hashes, Jenkins Hashes, etc. Not only should the ideal hash function provide independent and uniformly distributed results, it should also be as fast as possible to account for the mobile platforms' limited computation capabilities. Therefore, the final choice should preferably avoid widely used hashing algorithms, which generally run slow. After considering different factors, murmur32 was selected as the prime hash function for DiscoverFriends and the parameter to reduce the false positive probability was tuned to 2\%.

\textbf{Encryption} For symmetric key encryption, DiscoverFriends uses AES encryption.  In the network initialization phase, the secret key is the SHA-1 hash of the initiator's ID, trimmed to use only the first 128 bits. During subsequent message exchanges, the secret key is randomly generated using SHA1PRNG. Also, for this scheme, the cipher uses a PKCS5Padding transformation, which supports AES 128-bit keys.

As for asymmetric cryptography, a public/private key pair is generated using a RSA algorithm, in which key sizes are 1024 bits. With this key pair, DiscoverFriends generates self-signed X.509 certificates using the sun.security.x509 package. 

\textbf{Wi-Fi Broadcast} To implement Wi-Fi broadcast on Android, the IP address of the Wi-Fi interface is firstly retrieved. Then, the corresponding Wi-Fi broadcast address is derived. Afterwards, a UDP socket is established to send broadcast messages.


\subsection{Wi-Fi Direct}

First, a brief introduction of Wi-Fi Direct is necessary, followed
by a discussion on the choice and effectiveness of this technology.
The main purpose of Wi-Fi Direct is to enable peer-to-peer connections
without requiring a wireless access point (AP), allowing both one-to-one
and one-to-many connections among different devices through Wi-Fi
speeds. To enable compatibility for older devices using this newer
technology, only one device needs to be certified to the protocol
to allow for communication.
In addition, this technology enables
connection establishment with any device discovered within a 200 meter
radius, thereby leveraging the concept of locality. 

To introduce the reasoning behind choosing to utilize Wi-Fi Direct
technology over other methods, two general scenarios where DiscoverFriends
may be used are examined: a user is connected to a local area network
and a user is not connected to one. Lastly, this section ends with
an analysis of the applicability of the technology.

\subsubsection{Connected to LAN}

For the first scenario, the user is connected to a LAN, so we utilize
Android's Network Service Discovery to detect other devices that can
service DiscoverFriends. After detecting which devices are currently
running the application, a UDP broadcast is sent to these devices.
However, a few devices such as some HTC devices block these broadcasts,
so a workaround for these exception cases is needed. Here, Wi-Fi Direct
is not necessary because of the presence of a wireless network backbone. Additionally, helper nodes can be utilized to help to overcome the single hop limitation. The helper nodes would need to pass on  the initiator's Bloom filter for initial validation as well as any subsequent messages.

\subsubsection{Not Connected to LAN}

In a latter more probable scenario, the user enters a location where
there is no network infrastructure available. For example, the user,
Alice, goes to a park and wants to be able to communicate to all her
friends there. Here, she can use the Wi-Fi Direct technology in DiscoverFriends
to accomplish this goal through a two-step process.
\begin{enumerate}
\item The initiator discovers nearby devices and sets up the device, making
the user the group owner.
\item Attempts to establish a one-to-many connection with all the discovered
devices.
\end{enumerate}
For the purpose of this application, the group owner is set to be
the server and the connected devices to be the clients. Once connected,
the initiator is able to communicate to all its connected clients
using the channel managed by the WifiP2pManager. Wi-Fi Direct is suitable
for DiscoverFriends' assumptions because it leverages the locality
of this technology. However, all this assumes that the initiator,
as well as the friends, have the Wi-Fi Direct functionality. If the
friends do not support Wi-Fi Direct, then this technology is not suitable
for the DiscoverFriends environment. Although the initiator can set
up his phone as a wireless access point, the network name and passphrase
to connect to the device, which is generated by Android's WifiP2pManager,
are randomly generated. Also, note that how DiscoverFriends is designed,
discovery of new friends is not permitted unless the network is reinitialized.
This is not because of Wi-Fi Direct limitations but because of preventing
replay attacks. Therefore, it is not possible for the initiator to
provide his friends with the access information beforehand, thwarting
Wi-Fi Direct's usefulness in the application.

\subsubsection{Multi-Hop Capability}
\label{subsub:multi_hop_capability}

Wi-Fi Direct is the best option under the mobile ad hoc scenario, because of it makes multi-hop connection possible, which is another important merit and greatly extends the connectivity.
Android is well known for not supporting the Wi-Fi Ad-Hoc (IBSS) mode (the Issue 82 on the Android Open Source Project has been unsolved since 2008 \cite{adhoc.google}), so users cannot form ad hoc connection using the Wi-Fi without rooting their phones.

\begin{figure}[tb]
\includegraphics[width=1\columnwidth]{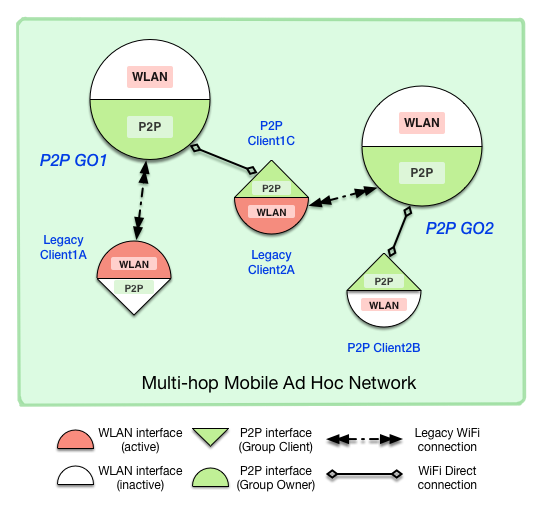}

\protect\caption{Wi-Fi Direct multi-hop connection topology}
\label{multihop-concept}
\end{figure}

Using Wi-Fi Direct to form multi-hop connection is conceptually straightforward.
The device which support Wi-Fi Direct comes with two Wi-Fi interfaces: one is the traditional Wi-Fi interface (legacy \textsf{wlan} interface) which is mainly used by regular WLAN connection, and the other is the Wi-Fi P2P interface (\textsf{p2p-p2p} or \textsf{p2p-wlan} interface) which is created for Wi-Fi Direct.
Thus, with both the legacy wlan interface and Wi-Fi P2P interface available, we can configure the intermediate nodes to be relay nodes by receiving traffic on one interface and sending it out on another.
The idea of multi-hop connection topology is shown in Figure \ref{multihop-concept}.
The node in the middle connects to Wi-Fi P2P Group Owner 1 (GO1) via its Wi-Fi P2P interface, and simultaneously connects to Wi-Fi P2P Group Owner 2 (GO2) via its traditional Wi-Fi interface.
Thus, message can be transmitted from GO1 to GO2 via the relay node, which forms a 2-hops connection.
Theoretically, the multi-hop connectivity is not limited, and can be extended to arbitrary number of nodes.

The logical equivalent of relaying the traffic on different interfaces is being able to configure routing table (or iptables, to be more accurate) in the runtime, which is easy enough for rooted devices.
The real challenge here is to implement the relay topology without root privilege, which is crucial for our application to be useful for the majority of the consumers without breaking Android’s security model.
Though it seems impossible for commodity-off-the-shelf unrooted Android phones, we can use Android's public API to make such topology work.
The key is Android's IPv6 support, which allow us to specify the network interface when sending out the packets, which perfectly solves connectivity issue.

Admittedly, there exists other workaround solutions to support multi-hop connection, for example, using the Bluetooth and Wi-Fi together \cite{firechat}. However, the Bluetooth is not even a competitive alternative for Wi-Fi Direct, in terms of data rate, transmission range and security.

\subsubsection{Applicability}

Finally, the applicability and general drawbacks of Wi-Fi Direct is
examined. Aside from compatibility with legacy devices, Wi-Fi Direct
draws significant power from the Android device, so it is not feasible
to keep Wi-Fi Direct on for an extended time \cite{k.-w._lim}, which
even the system warns the user when tries to enable the feature. However,
it is usable for DiscoverFriends because this application only runs
for a short period of time to exchange short messages. In addition,
the connection range is limited to 200 meters, but this is a suitable
distance for the application.

\subsection{OSN Adaptors}

There were many OSN adaptors to choose from, where the main ones were
Facebook, Google+, and Twitter. Here, DiscoverFriends implements the
first two. To support dynamic changes in friendship, DiscoverFriends needs only to add a new entry to the Bloom filter. The updated Bloom filter would then be sent to all peers.

\textbf{Facebook} Facebook was integrated in DiscoverFriends in order to obtain a user's
Facebook friends using Graph API v2.0. The Facebook IDs will then be placed into the initiator's request Bloom filter. Another Bloom filter is constructed by copying the previously constructed Bloom filter, which contains the hashed friends' IDs, and appending onto it, the user's own ID. Note the length of the friend list in the worst case for the expected insertions with a 2\% false positive probability, the optimal size of the Bloom filter is decided using Equation \ref{eq:k}. 



\textbf{Google+} The solution is extended to support Google+.  When using both OSNs, the two IDs representing a user is XORed to generate a new ID.  This new ID can only be identified by the initiator and the target
who is the initiator's friend on both OSNs. One issue is how to determine
which friends meet this requirement. Currently, the solution is to
hard code some common friends in the system. Later, it can be replaced
by a smarter mechanism such as comparing the name and email to infer
the potential common friend.


\section{Risk Analysis}

Risk analysis is done on three potential forms of network attack: replay attack, eavesdropping by a common friend, and traffic analysis of "Check-ins". For the replay and eavesdropping attacks, let us consider the simplified scenario where Alice is the initiator who connects to Bob. For the traffic analysis attack we consider two or more users sharing their "Check-ins" with multiple OSNs.

\subsection{Replay Attack}

In the case of a replay attack, we assume an adversary is able to
intercept and retransmit our application data. One purpose of this
attack may be to fraudulently establish a subsequent connection between
the adversary and the initiator\textquoteright s targeted group of
friends for a malicious rendezvous. Consider now, an adversary, Eve,
that is eavesdropping and wants to use that knowledge of the connection
to communicate to Bob by posing as Alice. In order for Eve to execute
a successful replay attack, she must firstly be authenticated by Bob.

For authentication, DiscoverFriends uses certificates which have
a limited validity period. Limited validity period refers to a successful instance of meeting up. Therefore, a new instance will result in requiring a new certificate in that respect such that the old certificate will be deem void, stripping away the possibility of a replay attack. So if Eve tries to replay data that contains an expired certificate, Bob will know. Procedure for certificate revocation
for the self-organized public-key management system is described in
\cite{k._sahadevaiah}. As a result, any attempts at replay attacks
are thwarted.

Although man-in-the-middle attacks are prominent in self-signed certificates
as a user does not have the certificate in advance for validation,
these malicious attempts are handled by proper detection of a compromised
node using the trust graph in the chosen key management scheme. In
other words, building a trust graph during the initial insecure network
initialization phase helps enable users to determine valid certificates
provided by the nodes contributing to the master graph. DiscoverFriends assumes that the social media site, initiator, and target are not compromised. This scheme's strong chain of trust also protects against Sybil attacks, where an
attacker can take on multiple identities in attempt to subvert the
reputation system of a peer-to-peer network. Therefore, self-signed
certificates by Eve will not be present in each node's CR, successfully
preventing man-in-the-middle attacks.

\subsection{Eavesdropping by a Common Friend}

When a common friend, Eve, eavesdrops, she is able to identify the
initiator because the hashed ID of one of his friends will come out
as a positive match. Without any additional security measurements,
Eve is able to decrypt subsequent messages sent from Alice using AES
with the matching ID from her friend list as the key. One such approach
is that instead of using the ID of the initiator as the key for AES,
the application uses the public key obtained from the certificate.
In the case of DiscoverFriends, an additional security measurement
is present as the AES key is encrypted using public-key cryptography.
As a result, the communication between initiator and target will be
private, preventing future eavesdropping.

It is also important to note that aside from common friends, bystanders
may be able to guess who the initiator is. This leads to guessing
the respective ID as it may be mnemonic such as a variation of the
initiator's name. Similar to the case of the common friend, the extra
security measurement of using the public key is necessary to hide
the conversation.

\subsection{OSN De-Anonymizaton Attack}

Suppose an OSN which operates a server wishes to de-anonymize a user and discover who posted a particular "Check-in" message. The user transmits a share to each OSN. As the user establishes an authenticated connection, each OSN knows which shares a user trasnmits. However, in order for a single OSN to reconstruct the message, all shares are required. A single OSN will not be able to reconstruct the original message as long as there is one honest OSN that does not collude.

When two or more users post their "Check-in", the shares will be XORed at each OSN. The XOR sufficiently randomizes the bits such that when two or more users participate, it is not possible to determine which user chose a particular database index. Thus, DiscoverFriends ensures users are able to anonymously post their "Check-in" as long as there are two or more users participating and there is at least one honest server that does not collude. 

\section{Performance Evaluation}

\subsection{Experiment Setup}

To evaluate the performance of DiscoverFriends, we utilize Android 4.0+ devices. As mentioned earlier, although the application supports
legacy devices, using any versions older than 4.0 will undermine the
application's purpose as an Android randomly generated passphrase
is needed to be known to Wi-Fi Direct unsupported phones.

Here, the network, computational, and storage costs are compared for three systems:
DiscoverFriends' using hybrid encryption, DiscoverFriends using only
AES with Bloom filters, and a system using prearranged ABE policy
trees as the initial baseline. In the following two sections, it is shown that using Bloom
filters and a hybrid encryption technique performs the best over the
latter two systems.

\subsection{Multi-hop Network Cost}
We use four Android Nexus 7 tablets to evaluate the performance of multi-hop Wi-Fi Direct network.
We connect tablets through the traditional Wi-Fi interface and the Wi-Fi P2P interface in a linear chain, as illustrated in Section~\ref{subsub:multi_hop_capability}.
In general, a $n$-hops scenario consists of $n+1$ phones chained via $n$ Wi-Fi P2P groups.
We start from the $1$-hop scenario, where two tablets form a group.
Both tablets are Wi-Fi P2P Group Owners and they connect via the traditional Wi-Fi interface.
We use iPerf to test the performance of the multi-hop network.
For the iPerf client, we direct the test traffic to a socket on localhost to our application.
Then application on the source node tablet will read the data sent by iPerf and relay it to the destination node via the multi-hop network.
When the destination node receives the data, it will redirect it to the iPerf server running on its localhost.
For simplicity, we utilize UDP protocol to send the traffic.
We measure the network throughput and packet loss by varying offered load.
Together with throughput, packet loss can be utilized to estimate the maximum bandwidth.
To evaluate a $(n+1)$-hops scenario, we add a tablet to the beginning of the $n$-hops chain, and connect it to the first tablet in the $n$-hops chain using the same connection topology.

\begin{figure}[tb]
\includegraphics[width=1\columnwidth]{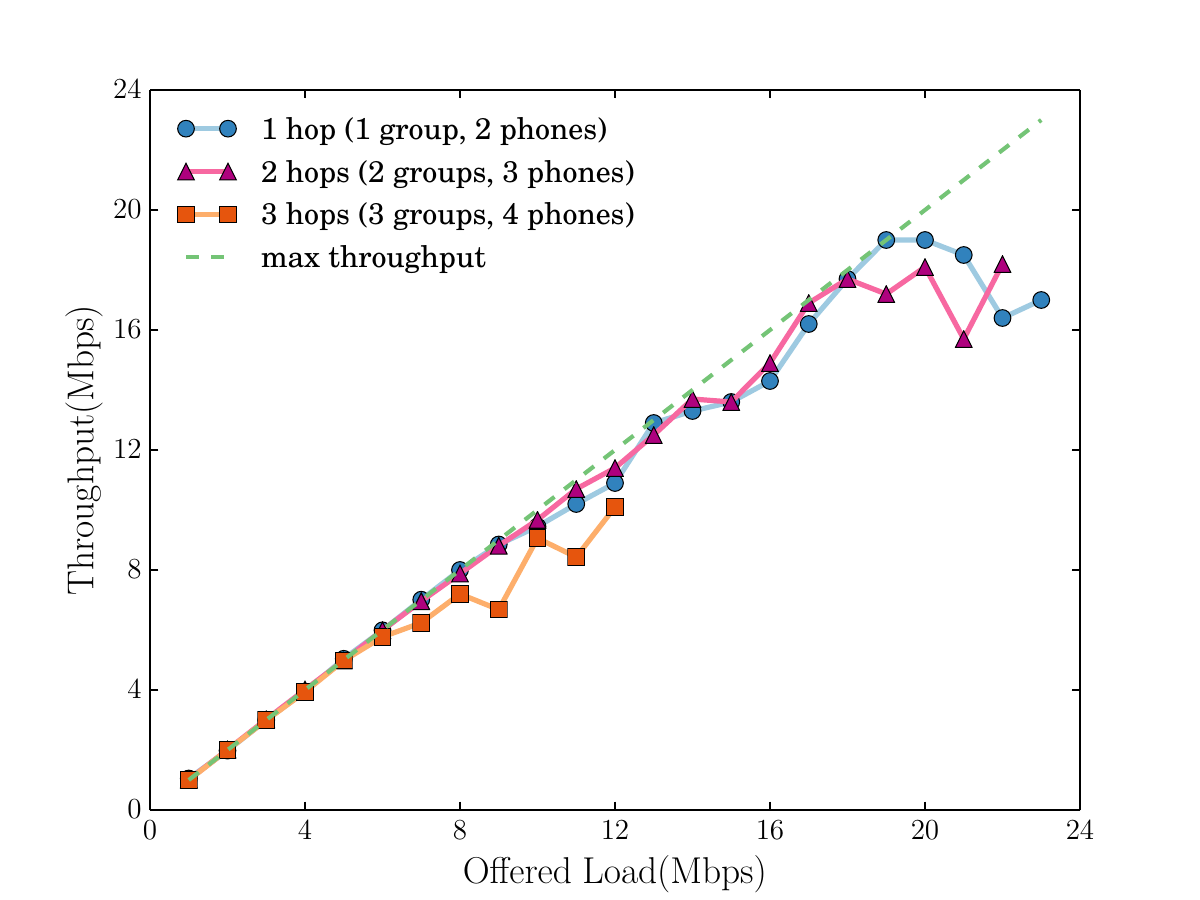}
\protect\caption{Wi-Fi Direct multi-hop throughput.}
\label{fig:multihop-throughput}
\end{figure}

Figure~\ref{fig:multihop-throughput} shows that as expected the throughput decreases with the number of hops. This degradation is due to the wireless interference between each hop as the chain utilizes the same channel in order to communicate per the Wi-Fi Direct standard. The maximum observed throughput is about $19$ Mbps for the $1$-hop scenario, about $18$ Mbps for the $2$-hops scenario, and about $10$ Mbps for the $3$-hops scenario. The severe degradation between the $2$-hops and $3$-hops is due to the spatial diversity and proximity placement of the tablets.

\begin{figure}[tb]
\includegraphics[width=1\columnwidth]{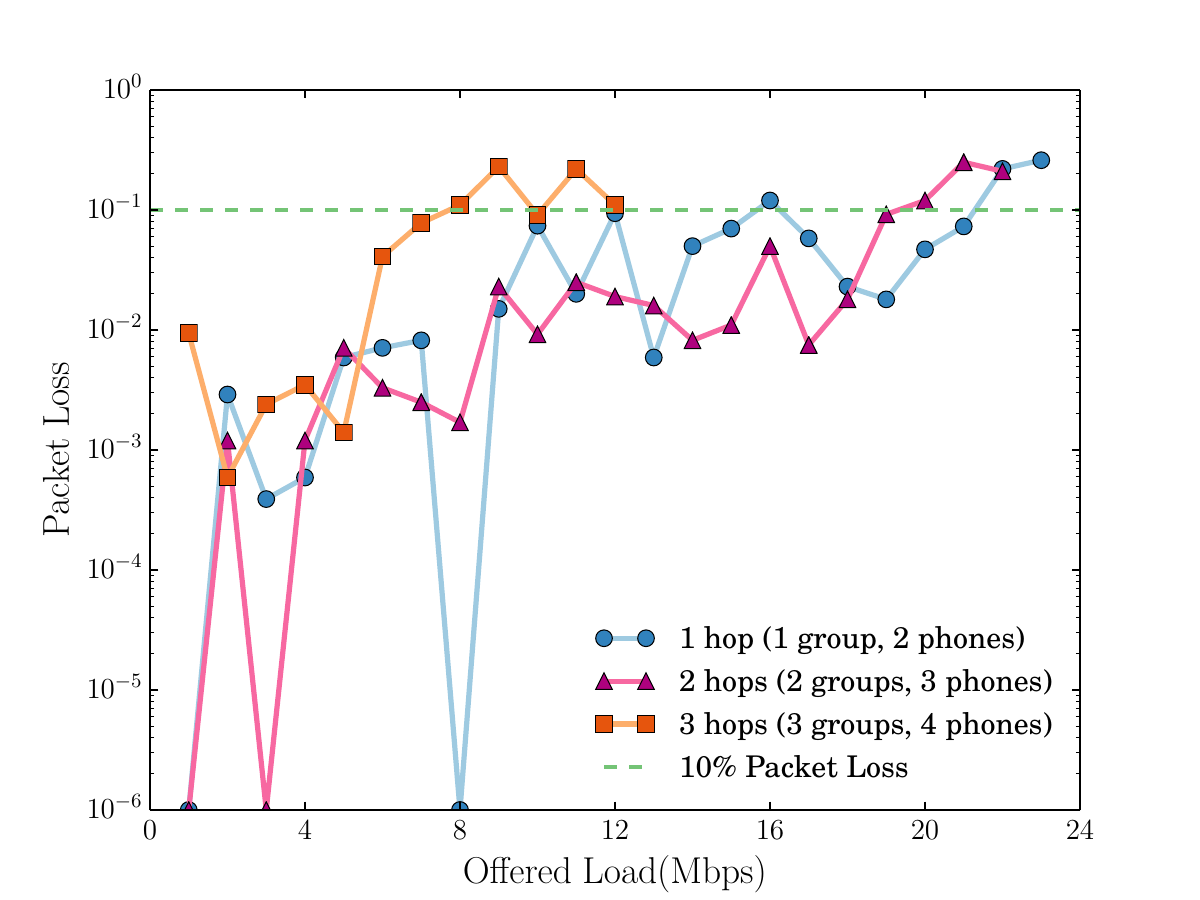}
\protect\caption{Wi-Fi Direct multi-hop packet loss.}
\label{fig:multihop-packetloss}
\end{figure}

Figure~\ref{fig:multihop-packetloss} depicts the packet loss as the number of hops are varied. As expected, the packet loss rapidly increases as we add load on the multi-hop network and first occurs at about $8$ Mbps. Both the $1$-hop and $2$-hops incur $10\%$ packet loss at around $22$ Mbps and $20$ Mbps respectively. The packet loss for the $3$-hops scenario is attributed to the wireless interference and tablet placement is with the throughput.


\subsection{Computational Cost}

Because encryption and decryption are computationally costly especially
in a MANET environment, energy-efficient approaches in applications
are necessary. The system using ABE keys has high computational cost
due to its large keys versus DiscoverFriends using only AES with Bloom
filters approach, which cuts down the cost to a fraction. However,
this approach trades off security for efficiency. Here, security suffers
due to by the small chance of false positives using Bloom filters,
whereas the advantage of using ABE is its stronger protection scheme.
The cost of encryption and key generation for ABE scales with the
number of attributes and the complexity of the access policy. Note
that performing offline key generation can speed up initial computational
cost while encryption must be done online. Therefore, a common approach
of using hybrid cryptosystems such as that in the final form of DiscoverFriends
lowers computational cost by only using the slow asymmetric algorithm
on keys rather than on messages while maintaining the necessary security
features.

As Function Secret Sharing (FSS) relies on the XOR rather than any expensive pairing operations, FSS is computationally efficient. Table \ref{fss_speed} evaluates the computational overhead for an anonymity set size of 2048 (we expect that users rarely exceed having 1000 friends). We evaluate messages up to 187 bytes, which supports our message size of 160 bytes. We have implemented FSS using Go. The server (acting as an OSN) is a Linux server running Linux 3.11.0 with a 2.2 GHz CPU with 24 cores and 32GB ram.

\begin{table}[tb]
\protect\caption{Function Secret Sharing Server Processing Time}
\label{fss_speed}

\centering{}%
\begin{tabular}{|c|c|}
\hline 
Message Size & Server Processing Time\tabularnewline
\hline 
62 bytes & 8 seconds \tabularnewline
\hline 
125 bytes & 14 seconds \tabularnewline
\hline 
187 bytes & 20 seconds \tabularnewline
\hline 
\end{tabular}
\end{table}

\subsection{Storage Cost}

The amount of storage used in DiscoverFriends is proportional to the
subset of successfully connected friends. It is shown that DiscoverFriends'
storage cost is significantly less than the other. In an experimentation,
two initiators who have 100 and 1,000 friends, respectively only want
to communicate to a subset of 10 friends. Table \ref{keystore_size}
depicts this experiment, and it is shown that DiscoverFriends' approach
requires less storage space compared to the other approach. The reasoning
behind this is because the Bloom filter prunes the initiator's full
friend list to a targeted group of friends. However, not everyone
in the list may be present or choose to respond to the initiator.
As a result, only the keys corresponding to the connected friends
are stored versus the entire set of ABE keys the other model has to
store. Because ABE generates an ASK for each user, the number of keys
corresponding to each user in an OSN may amount in the thousands,
thus it is not practical. Furthermore, under this approach, the social
network servers would know the keys and can defy attempts of covered
rendezvous. From this, the effectiveness of using Bloom filters come
to light as they only manage which users can decrypt the message using
a bit array structure while cutting OSN servers out of the picture.

While FSS is computationally efficient, this computational efficiency has a trade-off of high memory consumption. The space overhead is proportional to the anonymity set size desired in order to sufficiently anonymize and obfuscate the users' responses.  An anonymity set size of 2048 requires memory consumption of about 32GB ram.

\begin{table}[tb]
\protect\caption{KeyStore size comparison in KB}
\label{keystore_size}

\centering{}%
\begin{tabular}{|c|c|c|}
\hline 
 & 100 friends & 1000 friends\tabularnewline
\hline 
Use of Bloom Filters & 4.52 KB & 4.52 KB\tabularnewline
\hline 
No Use of Bloom Filters (e.g. ABE) & 44.9 KB & 449 KB\tabularnewline
\hline 
\end{tabular}
\end{table}

Another measurement was done on the sizes of DiscoverFriends' different
types of network packets: setup, certificate update, and normal. In
detail, the setup packet consists of the user's certificate and two
Bloom filters, where the parameters for the number of expected insertions
$n$ is 1000 and the false positive probability $p$ is 0.02. The
normal message has a maximum limit of 160 characters, which gets encrypted
using AES-128. These measurements are shown in Table \ref{packet_size}.

\begin{table}[tb]

\protect\caption{Total packet sizes in bytes}
\label{packet_size}

\centering{}%
\begin{tabular}{|c|c|}
\hline 
 & Total packet size\tabularnewline
\hline 
Setup & 2,516.59 bytes\tabularnewline
\hline 
FSS Share & 1,468,006 bytes\tabularnewline
\hline 
Certificate Update & 481 bytes\tabularnewline
\hline 
Normal & 176 bytes\tabularnewline
\hline 
\end{tabular}
\end{table}

\section{Related Work}

\subsection{Encryption Mechanisms}

Attribute-based encryption (ABE) is an alternative method of discovering
friends in place of Bloom filters. Users of ABE create three different
types of keys: ABE public key (APK), ABE master secret key (AMSK),
and ABE secret key (ASK). The first two are generated on a user bases
while the third is created per friend. The uses of these keys are
summarized as follows: APK is for encryption, AMSK is for secret keys
generation, ASK is for associating the users set of attributes. There
are two types of ABE schemes: Ciphertext-Policy and Key-Policy. The
Ciphertext Policy Attribute-Based Encryption (CP-ABE) consists of
four steps: Setup, Encrypt, KeyGen, and Decrypt. Similar to Bloom
filters, ABE targets a specific group that can decode the encrypted
message. Through this scheme, only a subset of all users with attributes
that match the access policy can decrypt the messages. Synchronization
is enabled using a key chain mechanism. The second scheme, Key-Policy
Attribute-Based Encryption (KP-ABE) is similar to the first scheme,
consisting of the same four algorithms but differs on attribute association.
Rather than associating the attributes with the user, KP-ABE instead
associates with the plaintext message. In other words, the decryption
policy enables only those users who match the ciphertext attributes,
which are associated with a plaintext message. In addition, there
are different flavors of ABE systems such as token-based ABE (tk-ABE)
\cite{m._j._hinek}, which protects against key cloning.

\subsection{Multi-Hop Connectivity}

Multi-hop connectivity is well studied in the literatures \cite{infocom04lee, comm14nishyama}, in terms of throughput \cite{ton07ng}, latency \cite{jsac99maltz} and even TCP performance \cite{vt07tan} etc.
But Wi-Fi Direct based multi-hop connection setup is getting more attention only in the recent years.
D. Camps-Mur et al. first studied the single group Wi-Fi Direct network topology \cite{ieeewc13cm}, focusing on tethering the 3G network access to the group clients.
Opportunistic group formation is also investigated in the work by M. Conti et al. \cite{ifip13conti}.
The Wi-Fi Direct based content sharing is first brought out by T. Duong et al. in \cite{icce12duong}.
C. Cassetti et al. designed a connection backbone scheme based on Wi-Fi Direct \cite{DBLP:journals/corr/CasettiCPVDG14}, and focused on the content centric routing performance. It is yet another interesting scheme, which their solution is solely based on two key observations in Android: that broadcast IP packets sent by the GO are always sent through GO's P2P interface whereas unicast packet is invariably sent through the GO Wi-Fi interface. However, neither author claimed these observations hold on any Android platform (later than Android 4.4) nor we can repeat their results in Android 5.0 or later system.
C. Yao et al. also worked on using Wi-Fi Direct to implement the D2D multi-hop network \cite{mobihoc15yao}. They assume the network topology will change often, and their data forwarding scheme is based such assumption. However, such scheme may incur high latency and cannot garantee the data will be transmitted to the destination, which is not suitable for our considered scenario.

\subsection{Applications}

Unlike LBS \cite{k._p._puttaswamy}, DiscoverFriends does not need
to know the user's location with GPS. The application exploits the
broadcast nature of wireless communication to directly find nearby
friends without the location information. Also, the user does not
expose location information to any server in DiscoverFriends, which
completely avoids the potential privacy leak.

In year 2009, a social network application called Safebook \cite{l._a._cutillo}
was implemented. It adopted a decentralized architecture and capitalized
on the trust relationships that existed outside of social networks.
Like DiscoverFriends, it addresses the concern where an omniscient
service provider such as Facebook or LinkedIn can intercept and potentially
monitor interactions between OSN users, essentially nullifying certain
privacy policies. Unlike DiscoverFriends however, this application
is not based off a MANET environment.

In year 2012, an algorithm for social networking on OLSR MANET utilizing
Delayed Tolerant Network (DTN) \cite{t._sanguankotchakom} was implemented.
The core idea was to discover friends based on similar interest within
the user's neighborhood. The solution realized that the use of Cosine
Similarity as the similarity metric yielded the highest number of
similar interest matching. Unlike this approach, DiscoverFriends'
algorithm has the list of friends already and instead, focuses on
how to communicate to them securely assuming they are nearby.

Recently in 2014, an idea to build symmetric private information retrieval
(PIR) systems using encrypted Bloom filters was conceived \cite{m._marlinspike}.
Rather than putting the user's ID in the Bloom filter as in DiscoverFriends,
a RSA signature of each user is placed instead. A client wanting to
query a local Bloom filter constructs a blinded query using David
Chaum's blinded signature scheme, which then gets signed by the server
and passed back. Receiving this, the client proceeds to unblind the
query to reveal the server's RSA signature for the targeted client
and checks if it exists within its local Bloom filter. However, this
approach like DiscoverFriends will fail with a reasonably large user
base as the Bloom filter sizes will be too large and inefficient to
transmit.

The first peer-to-peer anonymous communication, DC-nets,  was proposed in 1988 \cite{DBLP:journals/joc/Chaum88}. However, DC-nets are not scalable and would require each user to transmit several hundred MBs of data to achieve a scheme similar to DiscoverFriends. More recently in 2015 Riposte \cite{DBLP:conf/sp/Corrigan-GibbsB15} demonstrated a practical system for anonymous communication. However, Riposte requires several days to complete processing users' reponses as it relies on a seed-homomorphic PRG and expensive elliptic curves.

\section{Conclusions}

In this paper, we presented DiscoverFriends which is designed to find nearby Online Social Network (OSN) friends without disclosing one's location information to OSN servers, eliminating the potential privacy issue in using the Location-based services (LBS) of OSNs. DiscoverFriends is to our knowledge the first implementaton of a multi-hop Wi-Fi Direct based solution.  The use of Bloom filters-based message exchange and a hybrid encryption system with a self-organized public-key management enables confidentiality and authentication while providing higher security by bypassing OSN servers. DiscoverFriends further obscures user ID information by combining multiple OSNs, effectively protecting messages from being identified by any single OSN. By leveraging Function Secret Sharing, users are able to anonymously share their "Check-ins" across multiple OSNs. Under this model, it successfully prevents the effectiveness of eavesdropping in wireless communications as described by the risk analysis under three security threat models. 

There are several interesting areas of future work, including

\begin{itemize}
    \item A more robust mult-hop routing protocol and routing implementation
    \item A coding scheme to handle collisions for the Function Secret Sharing scheme
    \item An efficient disruption detection mechanism for handling malformed requests
    \item A large scale evaluation of Function Secret Sharing
\end{itemize}

With the design and implementation of DiscoverFriends, we have demonstrated and implemented using Android Wi-Fi Direct how to locate nearby Online Social Network(OSN) friends without violating users' privacy and without disclosing the location information to the OSNs.

\bibliographystyle{abbrv}
\def\bibfont{\footnotesize}
\small
{\footnotesize
\bibliography{bare_conf}}

\end{document}